\documentclass[%
 reprint,
 superscriptaddress,
 amsmath,amssymb,
 aps,
prl,
]{revtex4-2}

\usepackage{graphicx}
\usepackage{dcolumn}
\usepackage{bm}
\usepackage{color}
\usepackage{xcolor}

\usepackage{siunitx}
\newcommand{\curlyI}{\mathcal{I}}
\newcommand{\curlyL}{\mathcal{L}}
\newcommand{\curlyP}{\mathcal{P}}
\newcommand{\curlyC}{\mathcal{C}}
\newcommand{\Gtwo}{ G^{(2)} }

\newcommand{\Gtwoi}{ G^{(2)}_{i} }
\newcommand{\Gonei}{ G^{(1)}_{i} }
\newcommand{\Gtwopopi}{ G_{\text{pop},i}^{(2)} }
\newcommand{\GtwoHOMi}{ G_{\text{HOM},i}^{(2)} }
\newcommand{\twophotmat}{\rho^{2 \gamma} }

\newcommand{\D}{\mathrm{d}}
\newcommand{\E}{\mathrm{e}}
\newcommand{\I}{\mathrm{i}}
\newcommand{\hc}{\text{h.c.}}
\newcommand{\ketbra}[2]{|#1\rangle\negthinspace\langle#2|}
\newcommand{\ket}[1]{|#1\rangle}

\newcommand{\colorA}{red}
\newcommand{\colorB}{green}
\newcommand{\colorC}{blue}

\newcommand{\colorb}{black}
\newcommand{\colorc}{black}

\newcommand{\Tab}[1]{Table #1}
\newcommand{\Fig}[1]{Fig. #1}
\newcommand{\FIG}[1]{Figure #1}
\newcommand{\Figs}[1]{Figs. #1}
\newcommand{\FIGs}[1]{Figures #1}

\newcommand{\Eq}[1]{Eq. #1}

\newcommand{\Sup}[1]{Sup. Mat. #1}

\newcommand{\LabelA}{I}
\newcommand{\LabelB}{II}
\newcommand{\LabelC}{III}
\newcommand{\LabelD}{IV}

\usepackage{enumerate}
\usepackage{enumitem}
\usepackage{verbatim}
\usepackage{hyperref}
\usepackage[nolist]{acronym}

\begin{document}

\preprint{APS/123-QED}

\title{Swing-up dynamics in quantum emitter cavity systems} 

\author{Nils Heinisch}
 \affiliation{Department of Physics and Center for Optoelectronics and Photonics Paderborn (CeOPP), Paderborn University, Warburger Strasse 100, 33098 Paderborn, Germany}
 
\author{Nikolas K\"{o}cher}%
\affiliation{Department of Physics and Center for Optoelectronics and Photonics Paderborn (CeOPP), Paderborn University, Warburger Strasse 100, 33098 Paderborn, Germany}%

\author{David Bauch}%
\affiliation{Department of Physics and Center for Optoelectronics and Photonics Paderborn (CeOPP), Paderborn University, Warburger Strasse 100, 33098 Paderborn, Germany}%

\author{Stefan Schumacher}
\affiliation{Department of Physics and Center for Optoelectronics and Photonics Paderborn (CeOPP), Paderborn University, Warburger Strasse 100, 33098 Paderborn, Germany}%
\affiliation{Wyant College of Optical Sciences, University of Arizona, Tucson, AZ 85721, USA}%
\affiliation{Institute for Photonic Quantum Systems (PhoQS), Paderborn University, 33098 Paderborn, Germany}

\date{\today}

\begin{abstract}
 In the SUPER scheme (Swing-UP of the quantum EmitteR population), excitation of a quantum emitter is achieved with two off-resonant, red-detuned laser pulses. This allows generation of high-quality single photons without the need of complex laser stray light suppression or careful spectral filtering. In the present work, we extend this promising method to quantum emitters, specifically semiconductor quantum dots, inside a resonant optical cavity. A significant advantage of the Super scheme is identified in that it eliminates re-excitation of the quantum emitter by suppressing photon emission during the excitation cycle. This, in turn, leads to almost ideal single photon purity, overcoming a major factor typically limiting the quality of photons generated with quantum dots in high quality cavities. We further find that for cavity-mediated biexciton emission of degenerate photon pairs the Super scheme leads to near-perfect biexciton initialization with very high values of polarization entanglement of the emitted photon pairs. 
 \end{abstract}

\maketitle


\emph{Introduction} -- 
Semiconductor quantum dots (QDs) have been intensely studied as sources of single photons and entangled photon pairs \cite{Ding_2016,Hanschke_2018,Schweickert_2018,Chen_2018,Huber_2018,Liu_2019,Wang_PRL2019,Huber_2017,Mueller_2014,Wang_2016,He_2019,Koong_2021,Jonas_2022,Schumacher2012,Jahnke_2012}. The most efficient generation of highest-quality photons \cite{Praschan_2022,IlesSmith_2017,Ramsay_2010,Huber_2013}, however, as well as their extraction as optical information carriers \cite{Somaschi_2016,Wei_2014}, remain challenges on the road to using QDs as on-demand photon sources in quantum information processing architectures \cite{Varnava_2008,Gong_2010,Jennewein_2011}. A number of different excitation and photon extraction strategies have been explored and demonstrated over the years \cite{Reindl_2019,Thomas_2021,Ardelt_2014,Jayakumar_2013,Stufler_2006,Bauch}, however, each of the different approaches typically comes with specific limitations or difficulties \cite{IlesSmith_2017,Ramsay_2010,Huber_2013,Heinze_2015,Jonas_2022}. To achieve spectral separation of the excitation lasers and the emitted photons, in the last few years, excitation using dichromatic pulses has moved into the spotlight \cite{He_2019,Koong_2021,Vannucci_2022}. The recently introduced SUPER (Swing-UP of the quantum EmitteR population) scheme follows a similar approach \cite{SUPER,SUPERinAction,SUPERinActionJoens}, avoiding from the outset typical problems brought about by near-resonant optical excitation. In the Super scheme, by use of two off-resonant red-detuned laser pulses, phonon scattering is minimized at low temperatures \cite{Quilter_2015,Roy_2012}, and spectral filtering for photon extraction is easily performed \cite{Gustin_2020,Somaschi_2016,Wei_2014}.

\begin{figure}[t] 
\centering
\includegraphics[width=1.0\columnwidth]{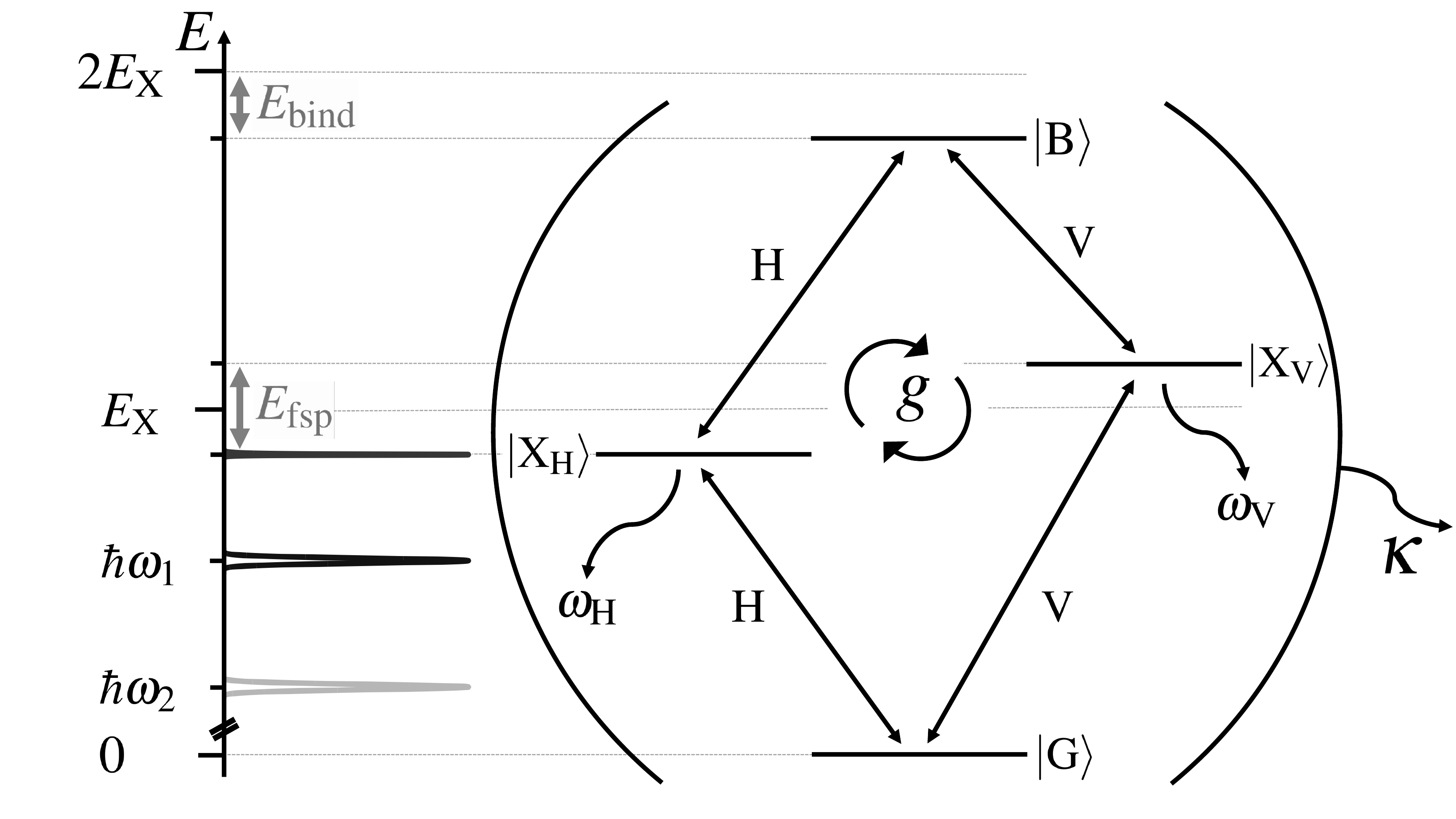} 
\caption{{\bf Schematics of quantum dot cavity system.} Electronic quantum dot states considered are ground state, $|\text{G}\rangle$, the two exciton states, $|\text{X}_\text{H}\rangle$ and $|\text{X}_\text{V}\rangle$, with fine structure splitting $E_{\text{fsp}}$, and the biexciton state, $|\text{B}\rangle$, with binding energy $E_{\text{bind}}$. The quantum dot transitions are coupled to the linearly polarized cavity modes with frequencies $\omega_\text{H}$ and $\omega_\text{V}$, with coupling strength $g$ and cavity photon loss $\kappa$. Spectra of the two red-detuned pulses of the Super scheme at $\hbar\omega_1$ and $\hbar\omega_2$ and exciton emission with typical widths are indicated. Energy differences shown are not to scale.}\label{Fig1}
\end{figure}

In the present work we extend this promising excitation method to a quantum emitter, specifically a semiconductor quantum dot, placed inside a resonant optical resonator as sketched in \Fig{\ref{Fig1}}. In that setup, for near-resonant optical excitation with a Gaussian laser pulse, cavity-accelerated photon emission (and then re-excitation) would occur already during the excitation cycle, spoiling the single-photon character of the emission \cite{Mueller_2015,Hanschke_2018}. For the Super excitation scheme, here we theoretically show that this emitter re-excitation is greatly suppressed via a laser-induced AC-Stark shift, leading to almost ideal single photon purity even for high quality cavities and approaching the strong coupling regime. This way the Super scheme allows us to overcome a major drawback that limits the quality of photons generated with quantum dot cavity systems. We further investigate the use of the Super scheme for the excitation of the biexciton state with subsequent cavity-mediated generation of pairs of degenerate polarization entangled photons \cite{Heinze_2017,Schumacher2012}. We show that in contrast to pulsed resonant two-photon excitation, the Super scheme avoids populating the cavity such that polarization entanglement is not reduced by the laser excitation \cite{Seidelmann_2022}.

\begin{figure}[t]
    \centering
    \includegraphics[width=1.0\columnwidth]{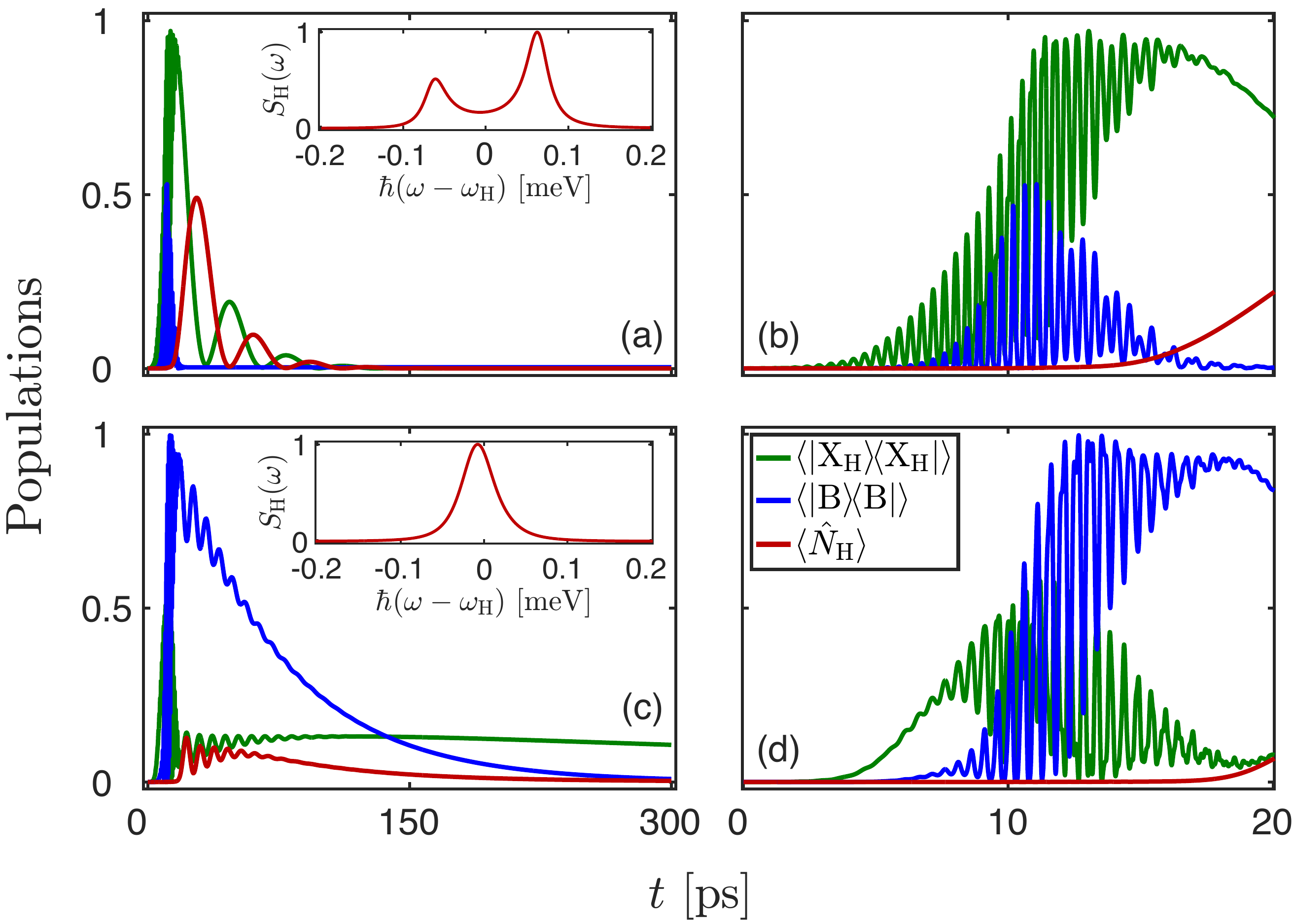}
    \caption{{\bf Electronic and photon populations for Super scheme excitation.} Shown are the H-exciton (green), biexciton (blue), and H-cavity (red) mode population dynamics. (a) and (b) show excitation of the H-exciton with the cavity tuned to the X$\to$G transitions for a QD with $E_\text{bind}=\SI{3}{\milli\eV}$. Laser pulse parameters as in first set of \Tab{\ref{Tab1}}. (c) and (d) show excitation of the biexciton with the cavity tuned to the two-photon B$\to$G transition with $E_\text{bind}=\SI{1}{\milli\eV}$. Laser pulse parameters as in fourth set given in \Tab{\ref{Tab1}}. While (a) and (c) show the time evolution up to $300\,\mathrm{ps}$, (b) and (d) only show the initial excitation period. The Gaussian laser pulses are about $3\,\mathrm{ps}$ long and centered at $10\,\mathrm{ps}$. Cavity coupling is $g=\SI{66}{\micro\eV}$ and cavity loss is $\hbar\kappa=g$. Insets in (a) and (c) show the normalized cavity H-mode emission spectra $S_\text{H}(\omega)$ in the relevant spectral range and with frequencies relative to the mode frequency $\omega_\text{H}$.}
    \label{Fig2}
\end{figure}

\emph{The Super scheme} -- 
We begin with a brief overview of the Super scheme as proposed in \cite{SUPER} and demonstrated in \cite{SUPERinAction,SUPERinActionJoens}. For a more detailed analysis see \cite{SUPERDressed}. For the excitation of the quantum emitter, the Super scheme uses a superposition of two off-resonant and red-detuned Gaussian laser pulses, resulting in the laser field amplitude
\begin{equation}\label{Field}
\begin{aligned}
\Omega(t) \ &= \ \frac{\Omega_1}{\sqrt{2 \pi \sigma_1^2}} \, \E^{-t^2/(2\sigma_1^2)} \, \E^{-\I \omega_1 t}\\[5pt]
    &\quad + \ \frac{\Omega_2}{\sqrt{2 \pi \sigma_2^2}} \, \E^{-(t-\Delta t)^2/(2\sigma_2^2)} \, \E^{-\I \omega_2 (t-\Delta t)+\I \phi}
\end{aligned}
\end{equation}
driving the emitter. 
Here, $\Omega_i$ denotes the pulse area, $\sigma_i$ the duration and $\omega_i$ the frequency of the respective pulse with $i=1,2$. $\Delta t$ denotes the temporal shift and $\phi$ the phase shift between the pulses ($\phi=0$ unless otherwise noted). When a quantum emitter (in the simplest case a two-level system) is excited from the ground state with an appropriate choice of pulse parameters, hereafter referred to as parameter sets, a Swing-UP of the quantum EmitteR population (SUPER) to the excited state is observed, initializing the emitter (cf. \Fig{\ref{Fig2}} for results with cavity). With both pulses being on the low-energy side of the resonant transition, at low temperatures, even in condensed matter environments, adverse phonon influences in the excitation process are minimized \cite{SUPER,Quilter_2015,Roy_2012}, and for sufficiently large pulse detunings spectral filtering of the photons emitted from the system is easily achieved.


\emph{The quantum dot cavity system} -- 
We model the lowest electronic excitations of a semiconductor QD as a diamond-shaped four-level system \cite{Hanschke_2018,Ota_2011}, as sketched in \Fig{\ref{Fig1}}, with typical model parameters for example for high quality InGaAs QDs \cite{Heinze_2015}. We note that the specific choice of system paramters does not influence the main results and conclusions of our study. We include the electronic ground state $\ket{\text{G}}$, two orthogonal exciton states $\ket{\text{X}_\text{H}}$ and $\ket{\text{X}_\text{V}}$, and the biexciton state $\ket{\text{B}}$. The two excitons have energies $E_{\text{X}_\text{H,V}}=E_\text{X} \mp \tfrac{E_\text{fsp}}{2}$ with exciton energy $E_\text{X} = \SI{1.366}{\eV}$ and fine structure splitting $E_\text{fsp}=\SI{2}{\micro\eV}$. We note that the exact choice of fine-structure splitting only has little influence on our excitation and single-photon emission results, however, for larger fine-structure splitting overall lower degrees of polarization entanglement are observed, see e.g. \cite{Heinze_2017}. The biexciton has a binding energy of $E_\text{bind}=\SI{3}{\milli\eV}$ unless stated otherwise. The electronic transitions are coupled to the respective orthogonal linearly polarized cavity modes with frequencies $\omega_i$ with $\ i=\text{H,V}$. Both cavity modes are coupled to the QD with coupling rate $g$ with photon loss $\kappa$. The laser pulses in \Eq{\eqref{Field}} drive the electronic transitions in the H-polarization channel. Starting from the system ground state, we compute the time evolution by explicitly solving the von-Neumann equation for the system density operator in Fock representation with external laser driving using a 4th order Runge-Kutta method with adaptive step size. Cavity losses are included in the Lindblad formalism. Further details on theoretical modelling and numerical implementation are given in \Sup{\LabelA}.

\begin{table}[t]
\centering
\begin{tabular}{c|c|c|c|c|c|c}
     $\hbar \Delta_1$ & $\hbar \Delta_2$ & $\Omega_1$ & $\Omega_2$ & $\sigma_1$ & $\sigma_2$ & $\Delta t$ \\
    $[\SI{}{\milli\eV}]$ & $[\SI{}{\milli\eV}]$ & $[\SI{}{\pi}]$ & $[\SI{}{\pi}]$ & $[\SI{}{\pico\s}]$ & $[\SI{}{\pico\s}]$ & $[\SI{}{\pico\s}]$\\\hline\hline
     {\color{\colorb}$-8$} & {\color{\colorb}$-17.5336$} & {\color{\colorb}$32.0007$} & {\color{\colorb}$32.0050$} & {\color{\colorb}$3.6058$} & {\color{\colorb}$3.4151$} & {\color{\colorb}$0.0138$} \\\hline
     {\color{\colorb}$-5$} & {\color{\colorb}$-11.3$} & {\color{\colorb}$25$} & {\color{\colorb}$33.33$} & {\color{\colorb}$4.0$} & {\color{\colorb}$4.0$} & {\color{\colorb}$0.0$} \\\hline\hline
     {\color{\colorc}$-5$} & {\color{\colorc}$-12.13567$} & {\color{\colorc}$30$} & {\color{\colorc}$30$} & {\color{\colorc}$1.3673$} & {\color{\colorc}$1.3673$} & {\color{\colorc}$0.0$}  \\\hline
     {\color{\colorc}$-5$} & {\color{\colorc}$-12.99$} & {\color{\colorc}$36.88$} & {\color{\colorc}$36.88$} & {\color{\colorc}$3.0$} & {\color{\colorc}$3.0$} & {\color{\colorc}$5.47$}  
\end{tabular}
\caption{{\bf Parameter sets for excitation.} The first two sets excite the H-exciton for a QD with $E_\text{bind}=\SI{3}{\milli\eV}$, the latter two sets excite the biexciton for a QD with $E_\text{bind}=\SI{1}{\milli\eV}$. Pulse detunings  are given relative to the G$\to$X transition energy, $\hbar\Delta_i=\hbar\omega_i-E_{\text{X}_\text{H}}$. Parameter combinations are not uniquely determined for exciton or biexciton excitation, respectively; we give two possible parameter sets for either case. Further discussion is given in the text.}
\label{Tab1}
\end{table}

\emph{Swing-up dynamics in four-level system with cavity} -- 
Based on the model detailed above, in this section we show that targeted excitation of one of the energy levels of the QD is possible and that the cavity mode does not generally hinder the Super scheme excitation. For the initialization of either exciton or biexciton state we study two different cases. In the first case the H-exciton is the target state for the excitation while the QD is in a cavity with modes resonant with the X$\to$G transitions with $\hbar\omega_i = E_i, \ i=\text{H,V}$. In the second case the biexciton state is the target state with the cavity resonant with the two-photon B$\to$G transition; $\hbar\omega_\text{H,V}=\tfrac{E_\text{B}}{2}$. For each of these scenarios we compare the results of the Super scheme excitation with the excitation with a (two-photon) resonant Gaussian pulse as a reference. 
In a first step we identify ideal excitation parameters (by simple parameter sweeps) such that near-unity population of the respective target states, H-exciton or biexciton, respectively, is achieved. Following the general rules for the Super scheme excitation \cite{SUPER,SUPERDressed}, for example the parameter sets listed in \Tab{\ref{Tab1}} fulfill these criteria. To emphasize the fact that parameter combinations are not uniquely determined for exciton or biexciton excitation, respectively, in \Tab{\ref{Tab1}} we give two possible parameter sets for either case; the first two sets excite the H-exciton, the last two sets excite the biexciton. We note that while virtually no inter-pulse phase dependence is observed for the first two and the last parameter set of \Tab{\ref{Tab1}}, the third parameter set is significantly phase dependent. Since achieving phase stability would be an additional experimental requirement, in the main text we only show results for the phase-insensitive parameter sets with zero phase difference between pulses, $\phi=\SI{0}{\pi}$; further discussion is given in \Sup{\LabelD}.

\begin{figure}[t] 
\centering
\includegraphics[width=1\columnwidth]{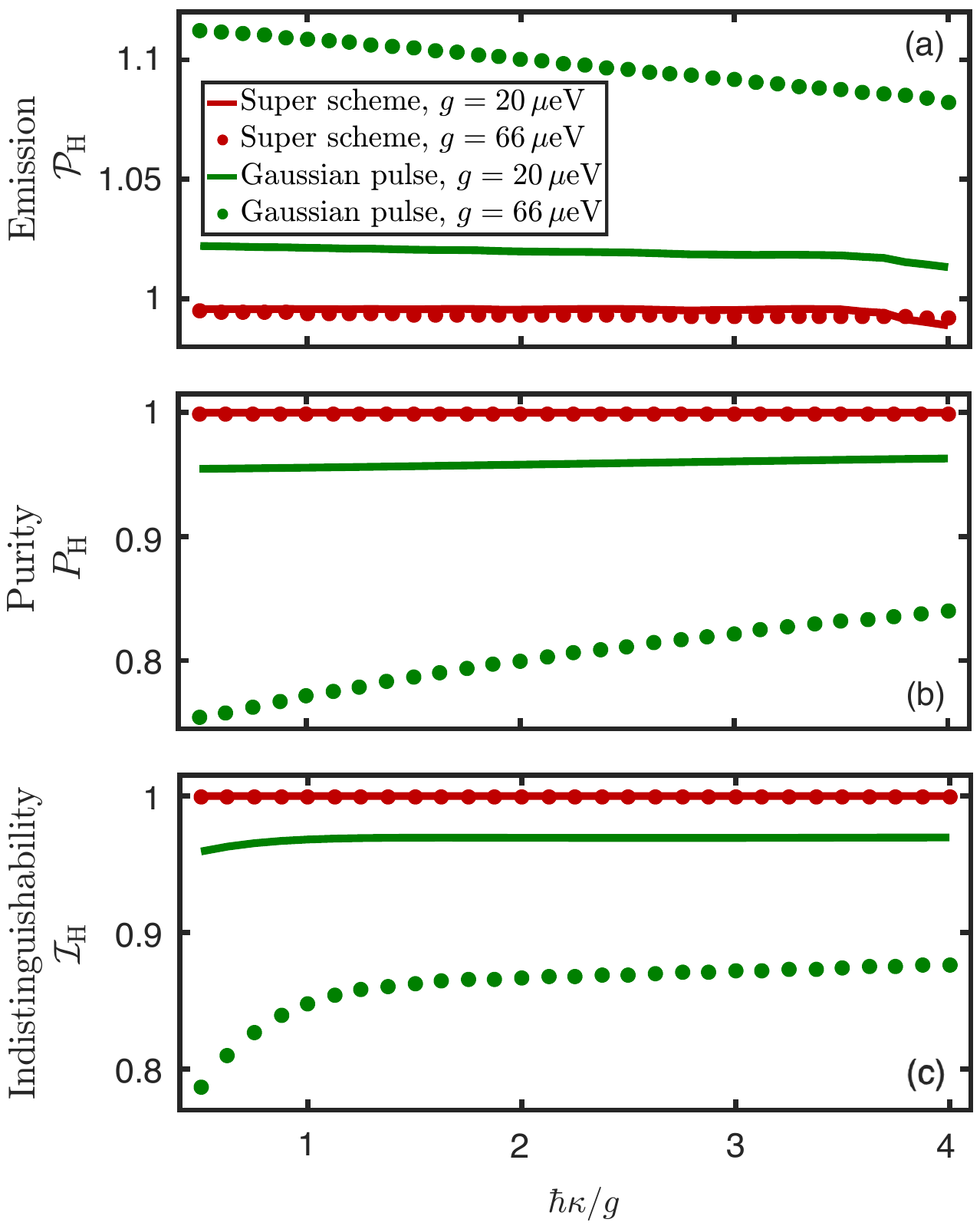} 
\caption{{\bf Single photon generation.} Quality measures for the photons emitted from the H-polarization cavity mode for a cavity resonant with the X$\to$G transition are shown. Excitation is with the Super scheme (red) or with a resonant Gaussian pulse (green), respectively, for two different QD-cavity couplings $g=\SI{20}{\micro\electronvolt}$ (solid lines) and  $g=\SI{66}{\micro\electronvolt}$ (dotted). Shown are (a) total emission probability $\curlyP_\text{H}$, (b) photon purity $P_\text{H}$, and (c) photon indistinguishability $\curlyI_\text{H}$. For the Super scheme the first parameter set of \Tab{\ref{Tab1}} is used while for the resonant Gaussian pulse a $\pi$-pulse with a width of $\sigma=\SI{5}{\pico\s}$ is used.}\label{Fig3}
\end{figure}

Now let us focus on the system excitation dynamics. \FIGs{\ref{Fig2}}(a) and (b) show the case, where we excite the H-exciton (first set from \Tab{\ref{Tab1}}), while \Figs{\ref{Fig2}}(c) and (d) show the targeted biexciton excitation (fourth set from \Tab{\ref{Tab1}}). The two pulses are centered around $\SI{10}{\pico\second}$ and about $\SI{3}{\pico\second}$ long; \Figs{\ref{Fig2}}(b) and (d) only show the main excitation window. Insets in \Figs{\ref{Fig2}(a)} and (c) show the cavity H-mode emission spectra. For this example, we use a cavity coupling strength of $g=\SI{66}{\micro\eV}$ along with $\hbar\kappa=g$. Results for lower cavity coupling, $\hbar\kappa=4g$, look qualitatively very similar as shown in \Fig{\ref{FigA1}} of \Sup{\LabelB}. The swing-up dynamics we observe is very similar to the observations without cavity \cite{SUPERDressed}. However, if we compare these swing-up results with the case of resonant excitation with a Gaussian $\pi$-pulse (not shown), one very important difference is that the cavity does not get populated with photons during the Super scheme excitation process, neither for exciton (\Fig{\ref{Fig2}}(b)) nor for biexciton (\Fig{\ref{Fig2}}(d)) excitation. For the exciton this is due to the AC-Stark shift \cite{Cosacchi_2020} induced by the strong, detuned pulses in the Super scheme. The QD transition is shifted out of resonance with the cavity and consequently no significant photon emission occurs during the excitation process (see \Sup{\LabelC} for more details). On the one hand, this is why the swing-up is very insensitive to the cavity parameters and not hindered by the presence of the cavity. In particular, there is no need to adjust the excitation parameters for different cavity parameters. On the other hand, premature emission and then emitter re-excitation, as observed for excitation with resonant pulses \cite{Mueller_2015,Hanschke_2018}, is greatly suppressed. Below we will show that this turns out to be one major advantage of the Super scheme as it leads to almost perfect single-photon purity even approaching strong emitter-cavity couplings. 

Next we turn our attention to the excitation of the biexciton. Using a two-photon resonant Gaussian pulse, a transfer of H-exciton excitation to the H-polarized cavity mode is observed during the excitation (not shown; see e.g. \cite{Bauch}). 
This causes a different emission behavior into the differently polarized modes, H or V, respectively, inducing which path information and reducing polarization entanglement. In contrast, below we show that for the Super scheme excitation, where photons are created equally in H- and V-cavity modes, for degenerate two-photon emission into the cavity mode, polarization entanglement is very insensitive to the excitation process.

The insets in \Fig{\ref{Fig2}} show the cavity emission spectra for the two cases of exciton (a) and biexciton (c) excitation. For exciton excitation a significant asymmetry is observed, again induced by the AC-Stark shift, consistent with \Sup{\LabelC}. 
For biexciton excitation no significant asymmetry is observable.


\emph{QD cavity system as single-photon source} -- 
Let us now investigate the potential of the Super scheme for excitation of a single-photon emitter. To most efficiently extract the emitted single photons in this case the cavity is resonant with the X$\to$G transition (corresponding to the case shown in \Figs{\ref{Fig2}}(a) and (b)). We now analyze the quality of the photons emitted from the H-cavity mode. Besides photon emission probability, $\curlyP_i$, we consider the purity $P_i$ as a measure for the single-photon character, and the photon indistinguishability, $\curlyI_i$, which measures how identical the generated photons are. Definitions of these quantities are given in \Sup{\LabelA}. Results are shown in \Fig{\ref{Fig3}} for different cavity loss rates $\hbar\kappa\in[\tfrac{1}{2}\,g,4\,g]$ and fixed cavity couplings $g=\SI{20}{\micro\eV}$ (solid lines) and $g=\SI{66}{\micro\eV}$ (dotted). In \Fig{\ref{Fig3}} we compare data for the emitter excited with the Super scheme (red; first parameter set of \Tab{\ref{Tab1}}) with data for resonant exciton excitation with a Gaussian $\pi$-pulse with width $\sigma=\SI{5}{\pico\s}$ (green). 
For excitation with the Gaussian pulse we consistently find emission probabilities larger than unity. This is a result of premature photon emission into the cavity mode while the pulse is still present and then re-excitation of the quantum emitter. This is most pronounced for small cavity loss $\kappa$ and large coupling $g$. For the Super scheme excitation, the emission probability is consistently just slightly below the ideal value of unity with an insignificant decline for increasing cavity loss $\kappa$. For excitation with the Gaussian pulse this decline is significantly more pronounced, especially for the larger cavity coupling $g=\SI{66}{\micro\eV}$. Apart from showing the insensitivity of the Super scheme to the cavity parameters, this indicates, that no re-excitation occurs for excitation with the Super scheme (in contrast to the resonant excitation). As already discussed above, for the Super scheme, this can be understood by the AC-Stark shift induced by the off-resonant pulses, shifting the X$\to$G transition out off resonance with the cavity mode during the excitation (cf. spectrally resolved emission shown in \Sup{\LabelC}). This is also reflected in the single-photon purity shown in \Fig{\ref{Fig3}}(b). For the Super scheme we find purity of almost exactly $1$ for both cavity coupling rates and all cavity loss rates investigated. When using resonant excitation, the purity increases with increasing cavity loss (more pronounced for larger cavity coupling $g$), however, it remains significantly below unity. This is consistent with emission probabilities in \Fig{\ref{Fig3}}(a) being larger than unity as a consequence of re-excitation events. 
The indistinguishability in \Fig{\ref{Fig3}}(c) shows similar trends as the purity. And again, the Super scheme convincingly outperforms the resonant excitation method, with near perfect indistinguishability values for both cavity couplings $g$ and all cavity losses $\kappa$. For completeness we note that we found very similar results for Super scheme excitation of a two-level system inside a resonant cavity (not shown). 

These results showcase the superiority of the Super scheme excitation (over resonant excitation) leading to cavity-enhanced emission of single-photons with near-ideal quality that are spectrally well separated from the excitation laser. We note that in the present study we focus on the intrinsic differences brought about by the different optical excitation methods studied, loss mechanisms that may largely differ for different quantum emitter systems and cavity designs such as dephasing, radiative loss, and phonon-induced cavity feeding are not discussed in detail here. 

\begin{figure}[t]
    \centering
    \includegraphics[width=1\columnwidth]{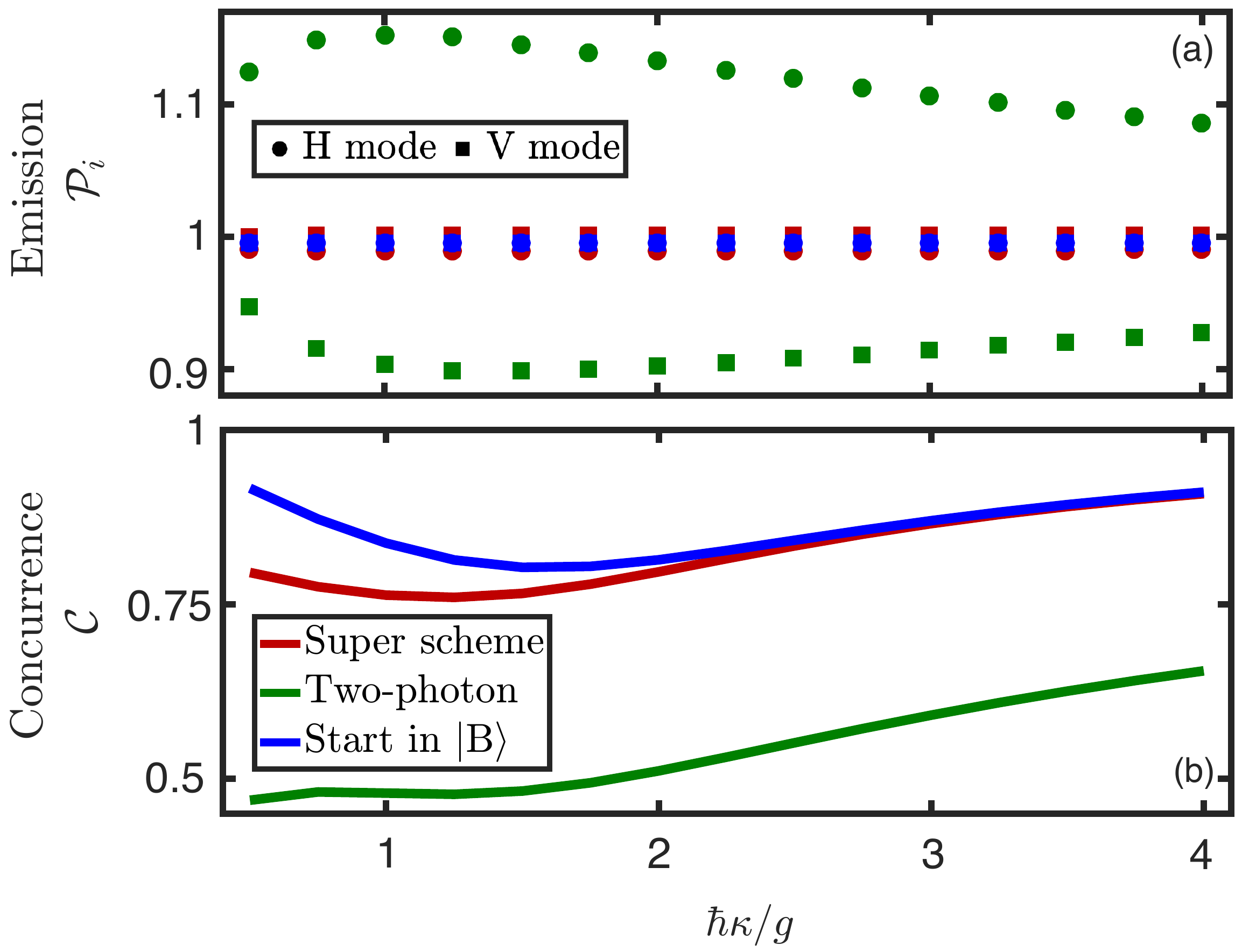}
    \caption{{\bf Polarization entangled photon-pairs.} Photon quality measures are shown for a cavity resonant with the degenerate B$\to$G two-photon transition for $g=\SI{66}{\micro\electronvolt}$. Shown are results for biexciton excitation with the Super scheme (\colorA), with a two-photon resonant Gaussian pulse (\colorB); $\hbar\omega=\tfrac{E_\text{B}}{2}$, and for an initially excited biexciton (\colorC). Shown are (a) the emission probabilities $\curlyP_\text{H,V}$ of the photons from the $\text{H}$- (points) and $\text{V}$- (squares) modes, respectively, and (b) the concurrence $\curlyC$. For the Super scheme the fourth parameter set of \Tab{\ref{Tab1}} is used, for the two-photon resonant Gaussian pulse a pulse area of $\Omega=\SI{3.3}{\pi}$ with a width of $\sigma=\SI{5}{\pico\s}$ is used.}
    \label{Fig4}
\end{figure} 

\emph{QD cavity system as source of entangled photon pairs} -- 
While Super scheme excitation of the quantum dot biexciton was discussed above, we now investigate the properties of the (polarization entangled) photons emitted into the cavity mode at half the biexciton energy in a resonant degenerate two-photon emission process \cite{Schumacher2012,Heinze_2017,Ota_2011}. To enhance this particular emission channel and emission probability $\curlyP_\text{H,V}$, a biexciton binding energy of $\SI{1}{\milli\electronvolt}$ and cavity coupling $g=\SI{66}{\micro\electronvolt}$ are used in this section. As a measure of polarization entanglement we calculate the concurrence, $\curlyC$. Definitions of $\curlyP_\text{H,V}$ and $\curlyC$ are given in \Sup{\LabelA}. \FIG{\ref{Fig4}} shows results for different cavity loss rates $\hbar\kappa\in[\tfrac{1}{2}\,g,4\,g]$. 
Besides the results for Super scheme excitation (\colorA; fourth parameter set of \Tab{\ref{Tab1}}), as references we include results for two-photon resonant biexciton excitation with a Gaussian pulse (\colorB; $\hbar\omega=\tfrac{E_\text{B}}{2}$, pulse area $\Omega=\SI{3.3}{\pi}$, width $\sigma=\SI{5}{\pico\s}$), and results for an initially excited biexciton (\colorC) without optical preparation. For the initially excited biexciton the emission probability equals unity for both H- and V-mode. For Super scheme excitation very similar values are obtained, with values for the H-mode just slightly below unity over the entire parameter range investigated. For two-photon resonant excitation, emission probabilities in the two cavity modes significantly differ, with values greater (less) than unity for the H-mode (V-mode) and pronounced changes with the cavity loss rate. In this case the excitation process leads to a residual population in the H-mode, significantly also reducing the observed degree of polarization entanglement as measured by the concurrence shown in \Fig{\ref{Fig4}}(b). We observe the highest concurrence values for the idealized scenario of the initially excited biexciton. However, this ideal case outperforms the Super scheme excitation only for the case of small cavity loss $\kappa$, with the Super scheme being a close second for larger $\kappa$. 

\emph{Conclusion} -- In the present work we have analyzed and demonstrated the potential that the Super scheme offers for the excitation of a photonic quantum emitter inside an optical cavity. Semiconductor quantum dots serve as a typical example here. While high excitation fidelities of the emmissive states (exciton and biexciton, respectively) are achieved for a wide range of parameters, quantum properties of the emitted single photons and polarization entangled pairs of photons are by far superior to the results found for resonant excitation with an optical $\pi$-pulse. As one main result, for single photon generation we find that photon emission and emitter re-excitation during the excitation cycle are entirely suppressed. This holds true even for systems approaching the strong coupling regime where photon emission occurs on timescales comparable to the duration of the excitation pulses. For Super scheme excitation, this leads to exceptionally high and near-ideal single-photon purity and indistinguishability over the entire parameter range investigated for our quantum emitter cavity system.

\begin{acknowledgments}
\emph{Acknowledgments} -- This work was supported by the Deutsche Forschungsgemeinschaft (DFG, German Research Foundation) through the transregional collaborative research center TRR142/3-2022 (231447078, project C09) and by the Paderborn Center for Parallel Computing, PC$^2$. 
\end{acknowledgments}

\bibliography{Paper}

\newpage


\section{Supplemental Material}
{\bf \LabelA $\ $ Theory} -- In this section, we present the theory used for our investigations. The Super scheme \cite{SUPER,SUPERDressed,SUPERinAction,SUPERinActionJoens} uses a superposition of two Gaussian pulses;
\begin{equation}\label{FieldAppendix}
\begin{aligned}
\Omega(t) \ &= \ \frac{\Omega_1}{\sqrt{2 \pi \sigma_1^2}} \, \E^{-t^2/(2\sigma_1^2)} \, \E^{-\I \omega_1 t}\\[5pt]
    &\quad + \ \frac{\Omega_2}{\sqrt{2 \pi \sigma_2^2}} \, \E^{-(t-\Delta t)^2/(2\sigma_2^2)} \, \E^{-\I \omega_2 (t-\Delta t)+\I \phi}.
\end{aligned}
\end{equation}
Here $\Omega_i$ for $i=1,2$ denotes the pulse area, $\sigma_i$ the duration and $\omega_i$ the frequency of the respective pulse. $\Delta t$ denotes the time shift and $\phi$ the phase shift between the two pulses. For legibility, we do not introduce the overall time shift for both pulses explicitly. However, for each choice of pulses, this shift must be introduced so that the pulses occur only after the time evolution begins in $t=0$. The full Hamiltonian in the lab frame is given by
\begin{equation}
    H \ = \ H_\text{QD} \ + \ H_\text{cav} \ + \ H_\text{QD-cav} \ + \ H_\text{QD-light},
\end{equation}
containing the free evolution of the QD \cite{Ota_2011},
\begin{equation}
    H_\text{QD} \ = \ \sum_{i=\text{G},\text{X}_\text{H,V},\text{B}} E_i \ketbra{i}{i},
\end{equation}    
with the respective QD energies $E_i,\ i=\text{G},\text{X}_\text{H,V},\text{B}$, the free evolution of the cavity,
\begin{equation}
    H_\text{cav} \ = \ \hbar \sum_{i=\text{H,V}} \omega_{i} a_i^\dagger a_i,
\end{equation}  
with cavity mode ladder operators $a_i^{(\dagger)}$ and frequencies $\omega_i,\ i=\text{H,V}$, the interaction Hamiltonian between QD and cavity,
\begin{equation}    
    H_\text{QD-cav} \ = \ g \sum_{i=\text{H,V}} \big(a_i\ketbra{\text{X}_i}{\text{G}} + a_i \ketbra{\text{B}}{\text{X}_i}\big) \ + \ \hc 
\end{equation}   
with real-valued coupling constant $g$ and the interaction Hamiltonian between QD and driving field,
\begin{equation}    
    H_\text{QD-light}  =  -\frac{\hbar}{2} \sum_{i=\text{H,V}} \Omega_i(t) \big( \ketbra{\text{X}_i}{\text{G}} + \ketbra{\text{B}}{\text{X}_i}\big)  +  \hc.
\end{equation}
For visualization of the model see \Fig{\ref{Fig1}} in the main part of this paper. Both interaction terms of the QD with either classical or quantized cavity light fields are treated in dipole and rotating wave approximation. The equation of motion for the system density operator \cite{Roy_2011} 
\begin{equation}\label{vNeqAppendix}
    \dot{\rho} \ = \ - \tfrac{\I}{\hbar} \, \big[ H , \rho \big] \ + \ \kappa \ \curlyL_\text{cav} \, \rho 
\end{equation}
contains the super operator 
\begin{equation}
    \curlyL_\text{cav} \, \rho = \sum_{i=\text{H,V}} \curlyL_{a_i} \, \rho,
\end{equation} 
describing cavity loss with the rate $\kappa$. It uses the common definition of the Lindblad operator for an operator $\hat{A}$,
\begin{equation}
\curlyL_{\hat{A}} \, \rho = \big( 2 \hat{A} \rho \hat{A}^\dagger - \hat{A}^\dagger \hat{A} \rho - \rho \hat{A}^\dagger \hat{A} \big).
\end{equation}
In all numeric investigations, we evolve \Eq{\eqref{vNeqAppendix}} in the interaction picture by performing transformations like $\tilde{\rho}= U^\dagger \rho U$, where $U=\exp \left[ - \tfrac{\I}{\hbar} (H_\text{QD} + H_\text{cav}) t \right]$. The properties of the cavity photons with polarization $i=\text{H,V}$ are the emission probability \cite{Gustin_2020},
\begin{equation}
    \curlyP_{i} \ = \ \kappa \int \D t \langle a_i^\dagger a_i \rangle(t),
\end{equation}
the indistinguishability \cite{Gustin_2018},
\begin{equation}
    \curlyI_i \ = \ 1 - \frac{\int \D t \int \D \tau \, 2\GtwoHOMi (t,\tau)}{\int \D t \int \D \tau \, \big( 2 \Gtwopopi (t,\tau) - |\langle a_i(t+\tau)\rangle\langle a_i^\dagger(t)\rangle |^2\big)},
\end{equation}
the purity (definition from \cite{SUPER}, and simplified using \cite{Gustin_2018} similar to $\curlyI_i$),
\begin{equation}
    P_i \ = \ 1 - p_i,
\end{equation}    
with
\begin{equation}    
    p_i \ = \ \frac{\int \D t \int \D \tau \, \Gtwoi (t,\tau)}{\int \D t \int \D \tau \, \Gtwopopi (t,\tau)},
\end{equation}
and the concurrence \cite{Wooters_1998,Cygorek_2018,Seidelmann_2020},
\begin{equation}
    \curlyC \ = \ \max \left\lbrace 0,\lambda_4-\lambda_3-\lambda_2-\lambda_1 \right\rbrace.
\end{equation}
Here, $\lambda_i, \, i=1,...,4$ are in rising order the eigenvalues of
\begin{equation}
    R= \sqrt{\sqrt{\twophotmat} \tilde{\rho}^{2\gamma} \sqrt{\twophotmat}},
\end{equation}
where
\begin{equation}
    \tilde{\rho}^{2\gamma} = \sigma_y \otimes (\twophotmat)^\ast \otimes \sigma_y,
\end{equation}
and $^\ast$ denotes the complex conjugate and with the two-photon density matrix
\begin{equation}\label{2PhotMatAppendix}
    \twophotmat_{ij} = \int \D t \int \D \tau \, \Gtwo_{ij} (t,\tau).
\end{equation}
In \Eq{\eqref{2PhotMatAppendix}} the indices $i,j=1,...4$ represent the two particle linear polarized basis in standard representation; $\left|1\right\rangle = \left|\text{H,H} \right\rangle$, $\left|2\right\rangle = \left|\text{H,V} \right\rangle$, $\left|3\right\rangle = \left|\text{V,H} \right\rangle$, $\left|4\right\rangle = \left|\text{V,V} \right\rangle$. In that representation, the two particle spin-flip matrix reads
\begin{equation}
    \sigma_y = \left(
 \begin{smallmatrix}
  & & & -1\\
  & & 1 & \\
  & 1 & & \\
  -1 & & & 
 \end{smallmatrix}
\right)
\end{equation}
and the index correspondence between matrix representation and $G^{(2)}$-functions uses $\left|i\right\rangle = \left|i_1,i_2 \right\rangle$ and reads 
\begin{equation}
    \Gtwo_{ij} (t,\tau) \ = \ \Gtwo_{i_1 i_2 j_1 j_2} (t,\tau).
\end{equation}
The miscellaneous $G^{(i)}$-functions are defined as
\begin{equation}
    \GtwoHOMi (t,\tau) \ = \ \tfrac{1}{2}\big( \Gtwopopi (t,\tau) + \Gtwoi (t,\tau) - |\Gonei (t,\tau)|^2  \big),
\end{equation}
\begin{equation}
    \Gtwopopi (t,\tau) \ = \ \langle a_i^\dagger a_i \rangle (t) \langle a_i^\dagger a_i \rangle (t+\tau),
\end{equation}
\begin{equation}
    \Gtwo_{ijkl} (t,\tau) \ = \ \langle a_i^\dagger (t) a_j^\dagger (t+\tau) a_k(t+\tau) a_l(t)\rangle,
\end{equation}
\begin{equation}    
    \Gtwo_{i} (t,\tau) \ = \ \Gtwo_{iiii} (t,\tau),
\end{equation}
\begin{equation}    
    \Gonei (t,\tau) \ = \ \langle a_i^\dagger (t) a_i(t+\tau) \rangle, 
\end{equation}
 where the indices run over the polarizations; $i,j,k,l=\text{H,V}$, again. For the evaluation of  the miscellaneous $G^{(i)}$-functions we use the quantum regression theorem (QRT) \cite{Carmichael_1999} 
\begin{equation}\label{QRTAppendix}
    \langle \hat{A}(t) \hat{B}(t+\tau) \hat{C}(t)\rangle \ = \ \operatorname{tr} \left\lbrace \Bar{\rho}(\tau)\hat{B} \right\rbrace,
\end{equation}
with
\begin{equation}
    \Bar{\rho}(\tau) = \left(\hat{C}\rho(t)\hat{A}\right)(\tau)
\end{equation}
for operators $\hat{A}$, $\hat{B}$ and $\hat{C}$. With a given time dependence the operator in the Heisenberg picture is referred to, without its counterpart in the Schr\"{o}dinger picture. These calculations are also handled in the interaction picture, for which the right-hand side in \Eq{\eqref{QRTAppendix}} further contains the mandatory transformation operators in appropriate places. As implemented by the QRT, the time evolution of the modified density operator $\Bar{\rho}$ is given by the same equation of motion as for $\rho$, \Eq{\eqref{vNeqAppendix}}. The cavity spectra are calculated using \cite{Mirza_2014}
\begin{equation}
    S_i(\omega)= \int \D t \int \D \tau \, \E^{-\I \omega \tau} \Gonei(t,\tau).
\end{equation}
Finally, we mention that the shorthand notation used above for double time integrals $\int \D t \int \D \tau$ replaces $\int_0^{t_\text{max}} \D t \int_0^{t_\text{max}-t} \D \tau$ where $t_\text{max}$ is chosen such that the system has fully decayed back to the ground state; $\ket{G,n_\text{H,V}=0}$. \\

\begin{figure}[h]
    \centering
    \includegraphics[width=1.0\columnwidth]{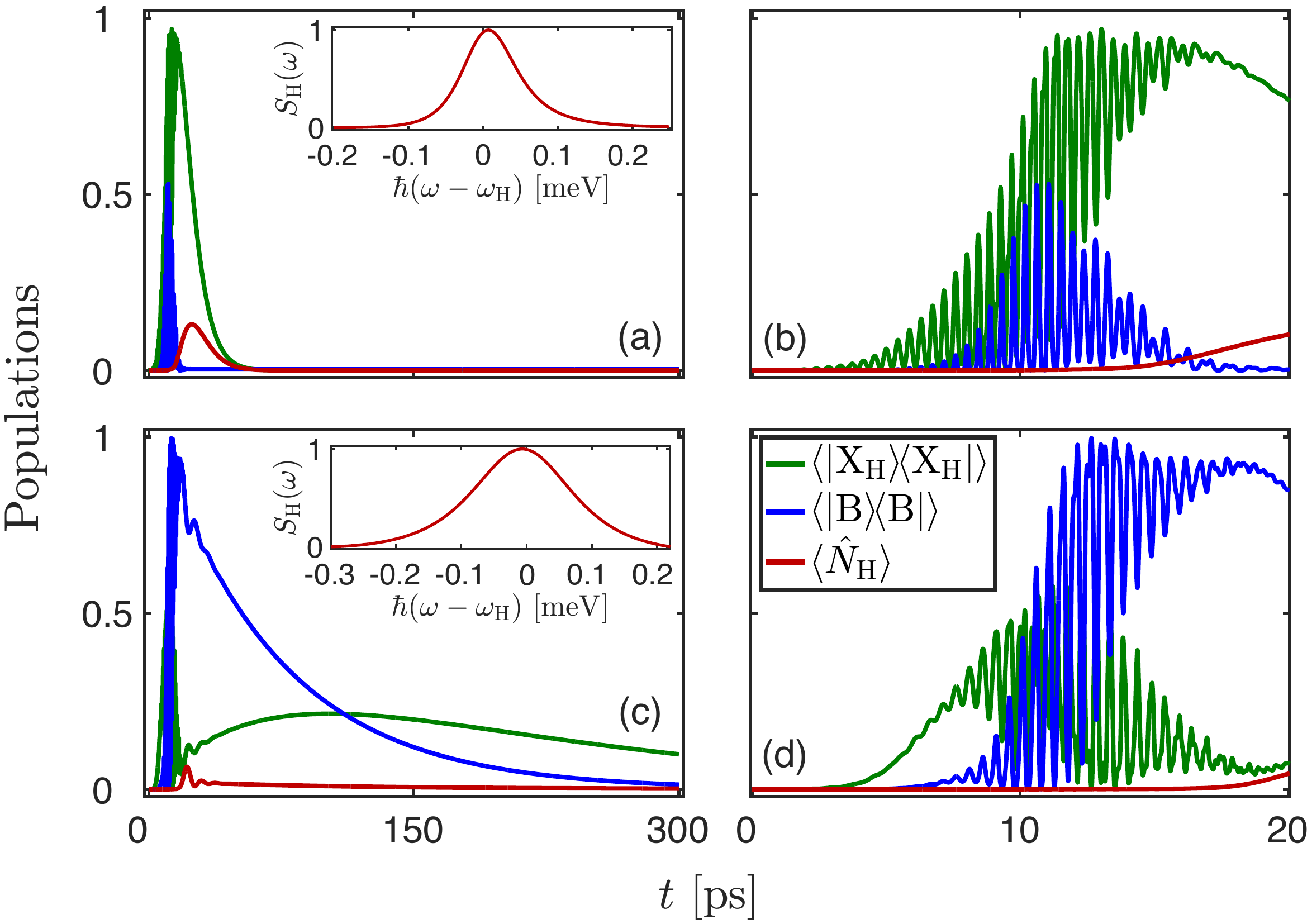}
    \caption{{\bf Electronic and photon populations for Super scheme excitation with larger cavity loss.} Shown are the H-exciton (green), biexciton (blue), and H-cavity (red) mode population dynamics. (a) and (b) show excitation of the H-exciton with the cavity tuned to the X$\to$G transitions for a QD with $E_\text{bind}=\SI{3}{\milli\eV}$. Laser pulse parameters as in first set of \Tab{\ref{Tab1}}. (c) and (d) show excitation of the biexciton with the cavity tuned to the two-photon B$\to$G transition with $E_\text{bind}=\SI{1}{\milli\eV}$. Laser pulse parameters as in fourth set given in \Tab{\ref{Tab1}}. While (a) and (c) show the time evolution up to $\SI{300}{\pico\second}$, (b) and (d) only show the initial excitation period. The Gaussian laser pulses are about $\SI{3}{\pico\second}$ long and centered at $\SI{10}{\pico\second}$. Cavity coupling is $g=\SI{66}{\micro\eV}$ and cavity loss is $\hbar\kappa=4g$. Insets in (a) and (c) show the normalized cavity H-mode emission spectra $S_\text{H}(\omega)$ with frequencies relative to the mode frequency $\omega_\text{H}$.}
    \label{FigA1}
\end{figure}

{\bf \LabelB $\ $ Details of the swing-up dynamics in a four-level system combined with a cavity} -- 
In this section, we give the details of our investigations on swing-up dynamics in QD cavity model systems as presented in the main part of this paper. There, we investigate whether excitation of one of the respective energy levels in the QD is possible with the Super scheme while the QD is coupled to a cavity. Concerning our investigations on excitation of the H-exciton here we give details of the time evolutions of the state populations given in \Figs{\ref{Fig2}(a)} and (b). During the excitation, we observe similar swing-up dynamics for the H-exciton as for the case without cavity; compare \cite{SUPERDressed}. Finally, the swing-up ends in an almost completely inverted system. While the H-mode is not populated during excitation, it performs the expected damped Rabi oscillations for the stronger cavity coupling together with the H-exciton population afterwards. Thus, the system reduces in the decay period approximately to the respective situation known from the Jaynes-Cummings model.
We show neither the V-exciton nor the V-mode population, as neither is populated. This is since the pulses have H-polarization. Further, none of the biexciton population that appears during the swing-up decays into the cavity, which is off-resonant w.r.t. the B$\to$X transitions. In \Figs{\ref{FigA1}(a)} and (b), we present the same situation but for a cavity with larger loss; $\hbar\kappa=4g$. The most significant difference from stronger coupling is the expected faster decay of the cavity population. Therefore, Rabi oscillations between the exciton and the cavity no longer occur. \\
Concerning our investigations on excitation of the biexciton we give now details of the time evolutions of the state populations given in \Figs{\ref{Fig2}(c)} and (d). It can be seen, that the swing-up excites the biexciton via the B-X-cascade. Also this is similar to the case without cavity. Compare again \cite{SUPERDressed}. As expected, the evolution of the V-polarized exciton and cavity mode show after the excitation similar behavior as their H- counterparts. During excitation they are zero (not shown). What is worth mentioning is that the mode populations behave the same also during excitation and rise just after successful biexciton preparation. After the cavity modes are populated, the expected decay sets in. In \Figs{\ref{FigA1}(c)} and (d), we present the same situation but for a cavity with larger loss; $\hbar\kappa=4g$. The most significant difference from the stronger coupling regime is the expected faster decay of the cavity population. Therefore, the population oscillations observable for $\hbar\kappa=g$ between the biexciton, exciton and cavity no longer occur. \\

\begin{figure}[h]
    \centering
    \includegraphics[width=1.0\columnwidth]{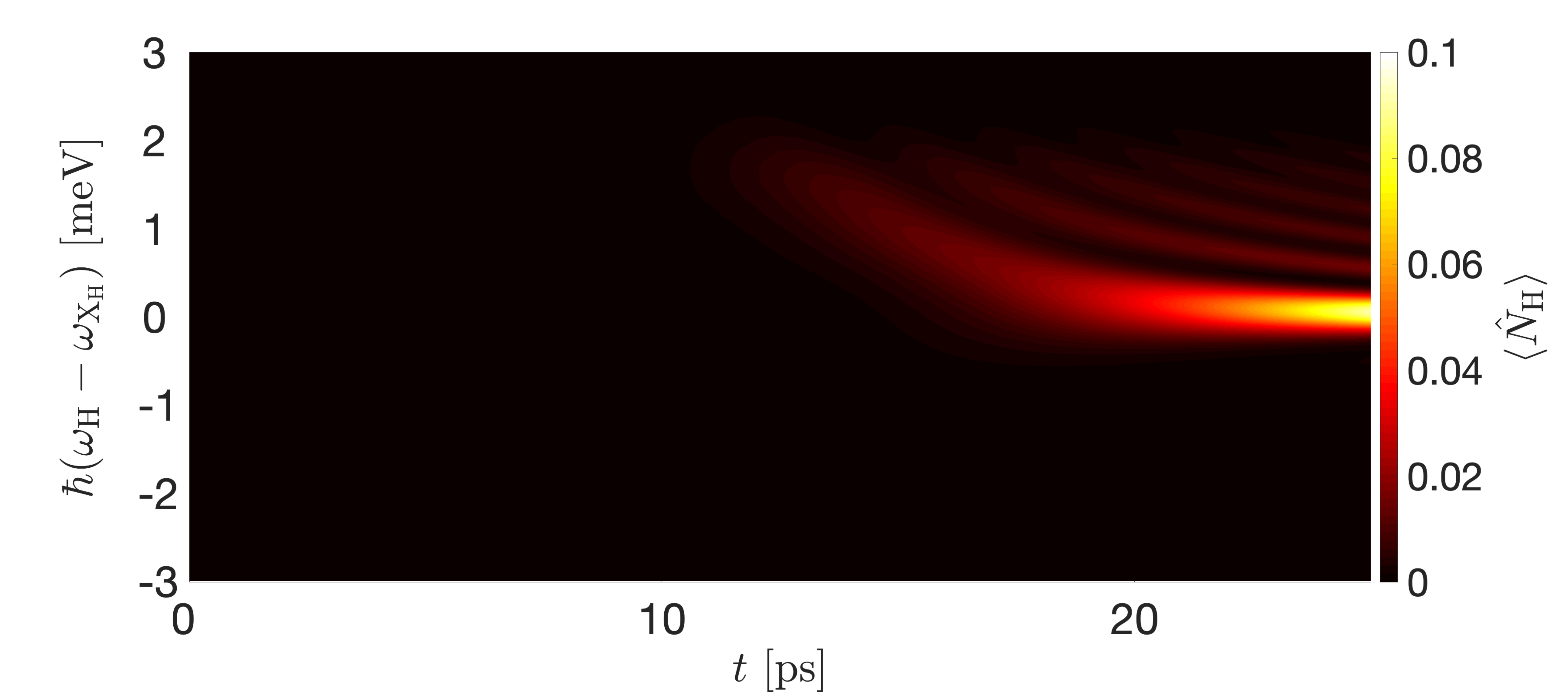}
    \caption{{\bf Cavity H-mode populations for different cavity frequencies.} Shown is the initial excitation period and the begin of the cavity enhanced decay with cavity frequencies $\omega_\text{H}$ relative to the X$\to$G transition frequency $\omega_{\text{X}_\text{H}}$. The QD has $E_\text{bind}=\SI{3}{\milli\eV}$. Laser pulse parameters as in first set of \Tab{\ref{Tab1}}. The Gaussian laser pulses are about $\SI{3}{\pico\second}$ long and centered at $\SI{10}{\pico\second}$. Cavity coupling is $g=\SI{66}{\micro\eV}$ and cavity loss is $\hbar\kappa=4g$.}
    \label{FigA2}
\end{figure}

\begin{figure}
\centering
\includegraphics[width=0.9\columnwidth]{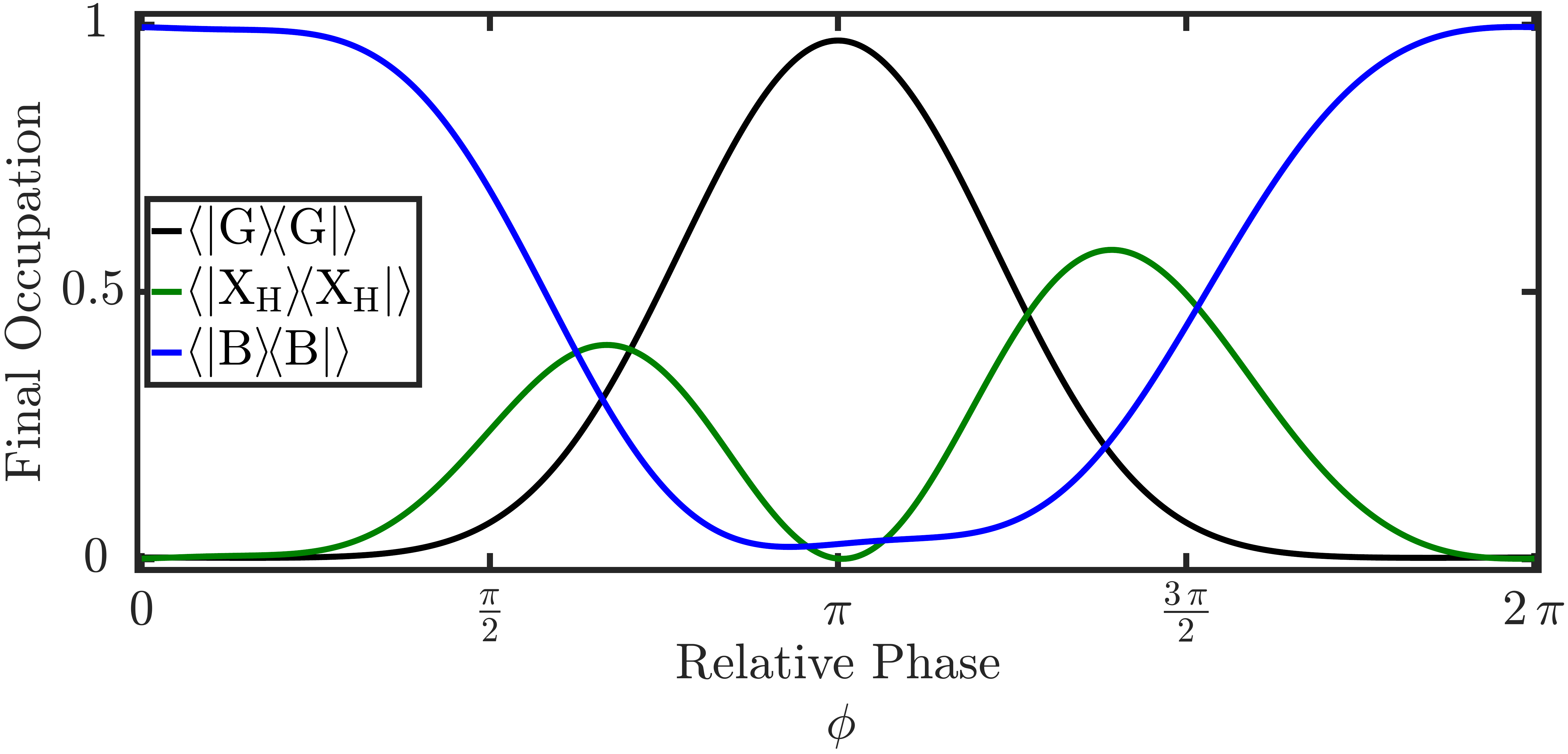}
\caption{{\bf Phase dependence of the Super scheme.} Analyzed are the populations of the QD following pulsed excitation without cavity interaction ($g=0$). Shown are the populations of the ground state, H-exciton and biexciton after pulse amplitudes have decayed to zero (final occupation). Laser pulses as in third set given in \Tab{\ref{Tab1}}, but for varying inter-pulse phase $\phi$. The QD has $E_\text{bind}=\SI{1}{\milli\eV}$.} \label{FigA3}
\end{figure}

{\bf \LabelC $\ $ Comment on light-field induced shifts of transition energies} -- That the AC-Stark effect indeed prevents re-excitation is confirmed by analyzing the cavity mode evolutions during the excitation for different cavity frequencies. \FIG{\ref{FigA2}} shows the cavity H-mode populations for different cavity frequencies during the excitation period and the beginning of the cavity enhanced decay. We reiterate, that the pulses are about $\SI{3}{\pico\second}$ long and centered at $\SI{10}{\pico\second}$. Clearly seen can be that the QD emits at shifted frequencies while the pulses are present. This shift takes its biggest value when the emission of photons in the cavity sets in and shrinks to zero with decreasing pulse amplitudes. This can be explained by the AC-Stark effect \cite{Cosacchi_2020}. The strong and detuned laser pulses shift the QD energies and therefore the energies at which photons are emitted. As the pulse amplitudes vary in time following their Gaussian envelopes, also the magnitude of the AC-Stark shift varies.
\\
{\bf \LabelD $\ $ Phase shift dependence of the Super scheme} -- In \cite{SUPER} and \cite{SUPERDressed}, only parameter sets independent of the phase shift $\phi$ between the two Gaussian pulses (we now simply call it phase or phase dependence) in \Eq{\eqref{FieldAppendix}} are presented. We also find parameter sets with no or negligible phase dependence, as stated in the main part of this paper. However, we discovered that such dependence does exist for some parameter sets. This is true for both the two-level system (TLS) and for the four-level system (FLS). See, for example, the third parameter set in \Tab{\ref{Tab1}} and the first two sets in \Tab{\ref{Tab2}}. \FIG{\ref{FigA3}} depicts a phase sweep for the third parameter set in \Tab{\ref{Tab1}}, allowing efficient biexciton excitation for the given phase of $\phi=0$. However, superpositions of populated exciton and biexciton can be achieved by changing the phase.
We observe similar behavior for the first two parameter sets in \Tab{\ref{Tab2}}. When we compare the parameter sets with and without phase dependence, we find the following difference for both model systems; TLS and FLS: in all cases with a phase dependence, the pulses are short in comparison to the time scale introduced by the detuning between the pulses; $\omega_\Delta =\omega_1 - \omega_2$. This observation is now compared to the previous work of \cite{SUPERDressed}, where the physics behind the Super scheme are explained using a dressed state approach. Because analytical comprehension is only provided for the TLS, an analysis of the phase dependence can only be provided for the TLS. Nonetheless, and in the light of the results we will present, the phase dependence in the FLS is most likely due to the same reasons. In summary, they use a dressed state approach to solve the TLS Hamiltonian for the first pulse. Then the second pulse is added, and the entire Super scheme is treated in the dressed state basis given by the first pulse. Finally, the Hamiltonian takes the form of the standard Rabi problem with a single pulse. Extending their findings, we discover that not only the interaction term, but also the free part, contains the phase in the resulting Hamiltonian. The phase has no significance in the interaction term, as it does in the standard Rabi problem. However, it may have in the free term. Following the notation of \cite{SUPERDressed}, the energies have a time dependence; $E_{\pm,2}=E_\pm \pm \hbar \Omega_2(t)c \tilde{c} \cos (\omega_\Delta t +\phi)$. Here, $c$ and $\tilde{c}$ give the coefficients for how the bare states form the dressed states. In this context, $\Omega_2(t)$ only refers to the envelope of the second pulse. For further details see \cite{SUPERDressed}. As noted there, for a sufficient excitation, a resonance condition similar to the standard Rabi problem must be satisfied; $\hbar \omega_\Delta \stackrel{!}{=} E_{+,2}-E_{-,2}= E_{+}-E_{-} + 2\hbar \Omega_2(t)c \tilde{c} \cos (\omega_\Delta t +\phi)$. Therefore, also the resonance frequency is time dependent and depends on the phase. For a sufficient number of oscillations of the cosine, the phase dependence averages out and only the mean energy difference can be considered. However, this is not true for pulse durations that are small compared to the time scale implemented by $\omega_\Delta$. This is consistent with our previous observations regarding the regions where the phase dependence occurs.
\begin{table}[h]
\centering
\begin{tabular}{c|c|c|c|c|c|c|c}
    $\hbar \Delta_1$ & $\hbar \Delta_2$ & $\Omega_1 $ & $\Omega_2 $ & $\sigma_1$ & $\sigma_2$ & $\Delta t $ & $\phi$ \\
    $ [\SI{}{\milli\eV}]$ & $ [\SI{}{\milli\eV}]$ & $ [\SI{}{\pi}]$ & $ [\SI{}{\pi}]$ & $ [\SI{}{\pico\s}]$ & $ [\SI{}{\pico\s}]$ & $[\SI{}{\pico\s}]$ & $ [\SI{}{\pi}]$ \\\hline\hline
    {\color{\colorc}$-8$} & {\color{\colorc}$-19$} & {\color{\colorc}$35$} & {\color{\colorc}$35$} & {\color{\colorc}$1.1897$} & {\color{\colorc}$1.2642$} & {\color{\colorc}$0$} & {\color{\colorc}$1.01$} \\\hline
    {\color{\colorc}$-5$} & {\color{\colorc}$-10.8688$} & {\color{\colorc}$32$} & {\color{\colorc}$29.6$} & {\color{\colorc}$1.5730$} & {\color{\colorc}$2.3195$} & {\color{\colorc}$0$} & {\color{\colorc}$0$} \\\hline
    {\color{\colorc}$-5$} & {\color{\colorc}$-11.8980$} & {\color{\colorc}$40$} & {\color{\colorc}$20.8081$} & {\color{\colorc}$3.0$} & {\color{\colorc}$3.0$} & {\color{\colorc}$0$} & {\color{\colorc}$0$}
\end{tabular}
\caption{{\bf Parameter sets for biexciton excitation for $E_\text{bind}=\SI{3}{\milli\eV}$.} Pulse detunings are given relative to the G$\to$X transition energy; $\hbar\Delta_i=\hbar\omega_i-E_{\text{X}_\text{H}}$.}
\label{Tab2}
\end{table}
\end{document}